# Determination of spin-orbit torques by thickness-dependent spin-orbit torque FMR measurement

Hongshi Li,[1,a)] Mahdi Jamali,[2] Delin Zhang,[2] Xuan Li,[2] and Jian-Ping Wang[2,b)]

[1] Department of Chemical Engineering and Materials Science, University of Minnesota, Minneapolis, MN 55455

[2] Department of Electrical and Computer Engineering, University of Minnesota, MN 55455

**Current induced spin-orbit torques (SOTs) in Fe/Pt bilayers have been investigated utilizing the spin-orbit torque ferromagnetic resonance (SOT-FMR) measurement. Characterization of thin films with different thicknesses indicates existence of a sizable field-like spin-orbit torque competing with the Oersted field induced torque (Oersted torque). The field-like torque is neglected in the standard SOT-FMR method and the presence of a strong field-like torque makes estimation of the spin Hall angle (SHA) problematic. Also, it is challenging to differentiate the field-like torque from the Oersted torque in a radiofrequency measurement. Based on the thickness dependence of field-like torque, anti-damping torque, and Oersted torque, the thickness-dependent SOT-FMR measurement is proposed as a more reliable, self-calibrated approach for characterization of spin-orbit torques.**

*Electronic address: a) hongshilih@163.com, b) jpwang@umn.edu

Spin-transfer torque (STT) has become an effective approach to electrically manipulate the magnetization and holds great promise for applications in memory and logic devices[1,2]. The spin polarized current required for STT operation has conventionally been generated from a ferromagnetic 'polarizer' layer in magnetic tunnel junction (MTJ) or spin valve[2]. Recently, it has been demonstrated both experimentally[3–9] and theoretically[10–13] that the spin torque can alternatively be provided by a nonmagnetic heavy metal (HM) layer adjacent to a ferromagnetic (FM) layer utilizing the spin-orbit coupling effect. This brings the possibility of spin current manipulation of the magnetization without exposing ferromagnets or MTJs to a large charge current[9]. The new approach exploits the coupling between electron spin and orbital motion to induce non-equilibrium spin accumulation and is therefore referred to as spin-orbit torques (SOTs).

Two commonly known mechanisms for SOTs are the spin Hall effect (SHE) which relies on the bulk properties of the heavy metal layer and the Rashba effect which takes place at the interface between FM and HM layers[13–15]. Meanwhile, there are two qualitatively different types of SOTs[13–15]: the anti-damping torque (ADT), $\tau_{AD} \propto \hat{m}\times(\hat{y}\times\hat{m})$, and the field-like torque (FLT), $\tau_{FL} \propto \hat{m}\times\hat{y}$, where the $\hat{m}$ is the magnetization unit vector and the $\hat{y}$ is the in-plane axis perpendicular to the current flow direction $\hat{x}$. While both the SHE and the Rashba effect, in principle, could give rise to the ADT and the FLT, the SHE is expected to contribute primarily to the ADT and Rashba effect is expected to be dominant on the FLT[13,15]. Despite the intensive experimental and theoretical studies on the SOTs and exciting progresses, SOT phenomena are still not well understood and require clarification of detailed mechanisms and even call for new models beyond the SHE and the Rashba effect.

Spin-orbit torque ferromagnetic resonance (SOT-FMR) based on the FM/NM bilayer structure has been proposed as a self-calibrated approach to characterize the spin Hall angle (SHA)[3]. The spin Hall angle $\theta_{SH}$ is defined as the ratio between the generated spin current $J_S$ and the injected charge current $J_C$. In the standard model for SOT-FMR, the ADT ($\tau_{AD}$) arising from the SHE and the Oersted field induced torque (Oersted torque, OT, $\tau_{Oe}$) are considered to be the only dominant mechanisms contributing to the measurement, and the FLT ($\tau_{FL}$) is not accounted for. The dynamics of the magnetization driven by SOTs could be described by the modified LLG equation[16,17]:

$$\frac{d\overrightarrow{M}}{dt} = -\gamma\overrightarrow{M}\times\overrightarrow{H}_{eff} + \alpha\overrightarrow{M}\times\frac{d\hat{m}}{dt} + \tau_{AD}\hat{m}\times(\hat{y}\times\hat{m}) - (\tau_{Oe}+\tau_{FL})\hat{m}\times\hat{y}, \tag{1}$$

assuming $\hat{x}$ is the current direction and $\hat{z}$ is perpendicular to the film. Here $\overrightarrow{M}$ is the total magnetization of the FM layer, $\hat{m}$ is the magnetization unit vector, $\overrightarrow{H}_{eff}$ is the sum of the external field and the anisotropy field, $\gamma$ is the gyromagnetic ratio, and $\alpha$ is the Gilbert damping constant. In the SOT-FMR measurement, a microwave current $I = I_{ac}\sin(\omega t)$ is injected into the FM/HM bilayer to induce the SOTs as well as the OT, which drive the magnetization of the FM layer to precess, as shown in Fig.1(a). The precession of magnetization results in an oscillatory resistance $R = R_0 + R_{ac}\sin(\omega t+\varphi)$ due to the anisotropic magnetoresistance (AMR) in FM layer. The injected microwave current mixed with the oscillatory resistance gives rise to a dc voltage component $V_{dc} = \frac{1}{2}I_{ac}R_{ac}\cos(\varphi)$. The $V_{dc}$ is plotted against the external field $H_{ext}$ applied during the measurement. At a fixed microwave frequency, a resonance peak will occur in the $V_{dc} - H_{ext}$ spectrum when $H_{ext}$ matches the resonance field $H_0$ for the ferromagnetic resonance (FMR), as displayed in Fig. 2(a).

The SOTs and the OT contribute differently to the symmetry of resonance peaks. In the case of in-plane anisotropy in the FM layer, the ADT lies in-plane and gives rise to a peak that is symmetric with respect to $H_0$. The FLT and OT are perpendicular to the film and give rise to a peak that is antisymmetric, as presented in Fig. 3(a). The SOT-FMR spectra thus can be fitted by combination of a symmetric and an antisymmetric Lorentzian function,

$$V = V_{Sym}\frac{\Delta H^2}{\Delta H^2 + (H - H_0)^2} + V_{Asym}\frac{\Delta H(H - H_0)}{\Delta H^2 + (H - H_0)^2}\ , \quad (2)$$

where $\Delta H$ is the linewidth of the Lorentzian functions, and $V_{Sym}$ and $V_{Asym}$ denote the strength of symmetric and antisymmetric components, respectively. The first term in Eq. (2) is symmetric and corresponds to the ADT, while the second term is antisymmetric and corresponds to the effective OT which is the combination of the OT and FLT. It should also be noted that only the Oersted field originated by the HM layer will contribute to the OT in the FM layer. The OT induced by the Oersted field generated within the FM layer cancels out when integrated over the FM layer thickness as long as the FM layer is reasonably uniform. Moreover, since the width of the thin films is orders of magnitude larger than the film thickness, the HM layer could be approximated as an infinitely large conducting plate. The Oersted field generated can thus be described by $H_{Oe} = J_C d_{HM} / 2$, which is constant in the perpendicular direction. As $\tau_{Oe} \propto H_{Oe} t_{FM} \propto J_C d_{HM} t_{FM}$, the strength of OT is proportional to both the FM and HM layer thickness. Here $d_{HM}$ and $t_{FM}$ are the thickness of FM and HM layer, respectively. In contrast, strengths of the ADT and the FLT are either independent of the FM and HM layer thickness or quickly saturate as the thickness increases[13–15] due to the short spin diffusion length of the Pt layer [3,18,19] and the short spin dephasing length of the Fe layer [20,21].

Two batches of samples with structures of Cr(3)/Pt(5)/Fe(2.5-10)/Ru(4) and Cr(3)/Pt(2.5-20)/Fe(5)/Ru(4) are deposited on thermally oxidized Si substrate by DC magnetron sputtering. The numbers in the brackets are layer thickness in nanometers. The Cr serves as the adhesion layer and the Ru is the capping layer. Ru have a relatively weak spin-orbit interaction compared to Pt and is expected to have only a minor contribution to SOTs. Meanwhile, as Ru has a similar resistivity as Cr, the additional Oersted field exerted on the Fe film by the Cr layer is largely compensated by that generated from Ru layer and makes no qualitative difference for the results in this research. Rectangular shaped microstrips with dimensions of $L(30\mu m)\times W(3-40\mu m)$ are patterned by optical lithography and Ar ion milling. Symmetric coplanar waveguides in the ground-signal-ground (GSG) form are utilized for microwave injection into the microstrips. An optical micrograph of the device is presented in Fig. 1(b). The symmetric GSG electrodes are utilized for microwave power injection in order to eliminate the unbalanced perpendicular Oersted field. The perpendicular field would  introduce an additional symmetric component to the SOT-FMR spectra[3] and complicate the measurement, as is confirmed in our previous experiments. A bias tee is used to inject microwave current and measure the resulting dc voltage at the same time. During the measurement, a microwave current with constant frequency is injected while a magnetic field is swept at an angle of $45^{\circ}$ with respect to the microstrips, and the output dc-voltage is measured at each magnetic field.

Fig. 2(a) shows the SOT-FMR spectra for a Pt(5)/Fe(10) thin film excited at microwave frequency between 8 and 18 GHz. Consistent with the FMR characteristic, the resonant field $H_0$ obtained from Eq. (2) in respect to the input microwave frequency $f$ has a quadratic behavior which agrees well with the Kittel formula $f=\frac{\gamma}{2\pi}\sqrt{(H_0+H_k)(H_0+H_k+4\pi M_{eff})}$ . The Kittel fitting for samples with different Fe thicknesses is presented in Fig. 2(b). Here, $\gamma$ is the gyromagnetic ratio, the $4\pi M_{eff}$ is the effective demagnetization field and the $H_k$ is the anisotropy field. The Kittel fitting indicates a $4\pi M_{eff}$ drop from 1.4 T for 10 nm thick Fe down to 0.97 T for 2.5 nm thick Fe, which might be explained by the existence of magnetic dead layers. SOT-FMR spectra measured with different microwave source powers at a fixed microwave frequency of 10 GHz are presented in Fig. 2(c). As expected, the peak position of the spectra is independent of the injected power. This is in agreement with Fig. 2(d), which shows that peak amplitude of both symmetric and antisymmetric components scale linearly with the input power, indicating the operation condition is in the linear regime. The dc-voltage due to the spin pumping and inverse spin Hall effect is also believed to be negligible as the amplitude of signal due to this side effect is typically one to two orders of magnitude smaller than that of the SOT-FMR[3,22].

In Fig. 3(a)-(c) one can see an interesting dependence of the antisymmetric component in the SOT-FMR spectra with respect to the Fe film thickness. At a fixed Pt thickness of 5 nm, as the Fe thickness decreases from 10 nm to 5 nm, the antisymmetric component decreases in amplitude by a factor of about 28. When the Fe thickness further decreases from 5 nm down to 2.5 nm the polarity of the antisymmetric component is altered while the amplitude increases by a factor of about 26. Meanwhile, the sign of symmetric component remains the same upon variation of the Fe layer thickness. Since the antisymmetric component of the SOT-FMR spectra corresponds to the effective OT, which is a combination of the OT and FLT, the sign reversal implies that the FLT and OT are competing with each other and have opposite signs. The reversal in sign of $\tau_{Oe} + \tau_{FL}$ by thickness variance could be explained by the dependence of OT on the magnetic film thickness, $\tau_{Oe} \propto J_c d_{HM} t_{FM}$. The OT is stronger than the FLT when the Fe layer is thick. In thinner Fe layers, however, the FLT dominates over the OT as its strength decreases linearly with the FM layer thickness. This phenomena could also be described quantitatively by the ratio of the effective OT over the ADT, $(\tau_{Oe} + \tau_{FL}) / \tau_{AD}$, which is plotted against the FM layer thickness in Fig. 3(d). As $t_{FM}$ is decreased from 10 nm to 2.5 nm the sign of the ratio changes from positive to negative and the existence of a FTL is clearly indicated by the negative intercept of the plotting line with the vertical axis.

The ratio of different torques could be determined by Eqn. (3), which is derived from the standard SOT-FMR model[3],

$$\frac{\tau_{Oe}+\tau_{FL}}{\tau_{AD}}=\frac{V_{Asym}}{V_{Sym}}[1+(4\pi M_{eff}/H_0)]^{-1/2}, \quad (3)$$

where $V_{Sym}$, $V_{Asym}$ and $H_0$ are the amplitudes of symmetric and asymmetric components in SOT-FMR spectrum, and resonance field under certain microwave frequency acquired by the Lorentzian curve fitting over experimental data, respectively. $4\pi M_{eff}$ can be estimated from Kittel fitting for each sample. At the same time, by differentiating between contribution from the FLT and the OT, the ratio $(\tau_{Oe}+\tau_{FL})/\tau_{AD}$ could be decomposed into the two terms in Eqn. (4),

$$\frac{\tau_{Oe}+\tau_{FL}}{\tau_{AD}}=(\frac{J_S}{J_C})^{-1}\frac{e\mu_0 M_s}{\hbar}t_{Fe}d_{Pt}+\frac{\tau_{FL}}{\tau_{AD}}. \quad (4)$$

The first term originates from the $\tau_{Oe}/\tau_{AD}$ ratio derived from the standard model[3] and has a linear dependence on both the FM and HM layer thickness. The second term is a constant that reflects the unknown FLT. The ratio $(\tau_{Oe}+\tau_{FL})/\tau_{AD}$ obtained from Eqn. (3) is plotted against the product of $t_{Fe}$ and $d_{Pt}$ in Fig. 4. Besides the first part of the experiment where Pt thickness is fixed at 5 nm and Fe thickness is varied from 2.5 to 10nm, we also perform a study on the effect of the Pt thickness on the SOTs. With the Fe layer thickness fixed at 10 nm, the Pt layer thickness is varied from 2.5 nm up to 20 nm. Fig.4 presents data from both batches of samples in the same plot. As seen, one single fitting line matches well for both sets of experimental data which is in accordance with Eqn. 4. This confirms our assumption that the strength of the ADT and the FLT is mostly independent of film thicknesses while that of the OT is proportional to film thicknesses.

Based on the discussion above, we propose utilizing the thickness-dependent study of SOT-

FMR characterization as a reliable technique to characterize the SOTs compared to the standard SOT-FMR method. One of the advantages of the thickness-dependent measurement is more reliable evaluation of SHA or the ratio between the ADT and OT. In the standard SOT-FMR model, the FLT is neglected, and as a result, the effective OT is mistaken as the OT. As the SHA is estimated from the ratio $\tau_{AD} / \tau_{Oe}$, this could result in overestimation of the SHA when the FLT is negative and underestimation when positive. Here the sign of the OT is defined as positive. The deviation could be rather significant when the film is relatively thin and the strength of the FLT is comparable to or larger than that of the OT. In contrast, the thickness-dependent measurement estimates the SHA more reliably by determining the slope of the $(\tau_{Oe} + \tau_{FL}) / \tau_{AD}$ ratio over film thickness in Fig. 4, as shown by Eqn.5,

$$\frac{J_S}{J_C} = \frac{e\mu_0 M_s}{\hbar} \times \frac{\tau_{AD}}{\tau_{Oe} / (t_{Fe} d_{Pt})}. \quad (5)$$

From linear curve fitting of Fig. 4, the $J_S / J_C$, or the SHA, in the Pt layer is estimated to be 0.20. Moreover, the thickness-dependent characterization has retained the self-calibrated nature of the SOT-FMR measurement. In other words, the spin-orbit torques and the corresponding charge current are measured relative to each other in a single SOT-FMR spectrum, rather than being measure separately, and there is thus no need to determine the resistance of each layer. This advantage makes a HM-layer-thickness-dependent SOT-FMR characterization of the SHA very useful for the cases where film structures are complicated or the proportion of charge current running through the HM layer is challenging to determine.

The second advantage of the thickness-dependent characterization is the quantitative determination of strength of the FLT. As charge current flowing into the HM layer could be tricky to determine accurately, it is challenging to differentiate the FLT from the OT in a single SOT-FMR spectrum. This is even more the case for radiofrequency measurement due to the current leaking caused by parasitic capacitances. This problem could be solved utilizing linear dependence of the OT on film thickness. In Fig. 4, the ratio $\tau_{FL} / \tau_{AD}$, which is equivalent to the intersection of the fitting line with the vertical axis, is fitted to be -0.30. This means that the strength of the FLT is comparable to that of the ADT in strength in our Fe/Pt bilayer structure and cannot be neglected in either property characterizations or device applications.

In conclusion, SOT-FMR measurement has been conducted in Fe/Pt bilayer with different film thicknesses of the Fe and Pt layers. Characterization of thin films with different thicknesses indicates existence of a sizable field-like spin-orbit torque competing with the Oersted torque. Utilizing the thickness dependences of different torques, the SHA is estimated to be 0.20 and the strength of the FLT relative to ADT is determined to be -0.30. Our result indicates that the FLT in our Fe/Pt bilayer structure is comparable to the ADT in strength and cannot be neglected. Based on this experiment, the thickness-dependent SOT-FMR measurement is proposed as an effective approach for the characterizations of SOTs compared to the standard SOT-FMR metond as it provides a more reliable estimation of the SHA. Moreover, it provides a feasible approach to detect and quantify the FLT neglected in the standard method.

Acknowledgement

This work was partially supported by the C-SPIN center, one of six STARnet program research centers, and National Science Foundation Nanoelectronics Beyond 2020 (Grant No.NSF NEB 1124831). Parts of this work were carried out in the Minnesota Nano Center which receives partial support from NSF through the NNIN program.

Figure Captions

FIG. 1. (a) Microwave current injected into the Fe/Pt bilayer induces the SOTs as well as the OT, which excite the precession of magnetization. For a FM layer with in-plane anisotropy, the ADT lays in plane of the film while the FLT and OT are perpendicular to the film. External field $H_{ext}$ is applied at an oblique angle to tilt the magnetization away from the easy axis and enhance the anisotropic magnetoresistance effect. (b) Schematic of the device utilized for the SOT-FMR measurement. The microstrip of Fe/Pt is contacted by symmetric GSG electrodes and a bias tee is used to inject microwave power and read the resulting dc voltage.

FIG. 2. (a) SOT-FMR spectra on a Pt(5)/Fe(10) thin film measured under different microwave frequencies between 8-18 GHz and with an external field applied at 45° relative to the microstrip. The numbers in brackets are thickness in nanometers. (b) Resonance frequency $f$ plotted against resonant field $H_0$ in Pt(5)/Fe(2.5, 5, 10) thin films. The solid lines are the results for Kittel fitting. (c) SOT-FMR spectra of a Pt(5)/Fe(2.5) thin film measured under an excitation frequency of 10 GHz and with microwave source powers varied from 400 mV to 2800 mV. (d) Amplitudes of the symmetric and antisymmetric components in the SOT-FMR spectra plotted against the injected microwave power.

FIG. 3. The SOT-FMR spectra for bilayer with structure of (a) Pt(5)/Fe(2.5), (b) Pt(5)/Fe(5) and (c) Pt(5)/Fe(10) measured at an excitation frequency of 10 GHz. The thicknesses in brackets are in nanometers. The experiment data (black) is overlaid with the fitted symmetric Lorentzian (red) and antisymmetric Lorentzian (blue) curve, as well as their summation (green). A clear reversal in sign of the antisymmetric component is observed as the thickness of Fe layer is varied. (d) The ratio $(\tau_{Oe} + \tau_{FL}) / \tau_{AD}$ plotted against the Fe layer thickness quantitatively shows the sign reversal of $\tau_{Oe} + \tau_{FL}$.

Fig. 4 The ratio $(\tau_{Oe} + \tau_{FL}) / \tau_{AD}$ for Fe/Pt bilayers is plotted against the product of $t_{Fe}$ and $d_{Pt}$. The red dots represent the ratios for bilayers with Pt layer thickness fixed at 5 nm and Fe layer thickness varied from 2.5-10 nm. The blue dots represent the ratios for bilayers with Fe layer thickness fixed at 10 nm and Pt layer thickness varied from 2.5-20 nm. The black curve is the outcome of linear fitting.

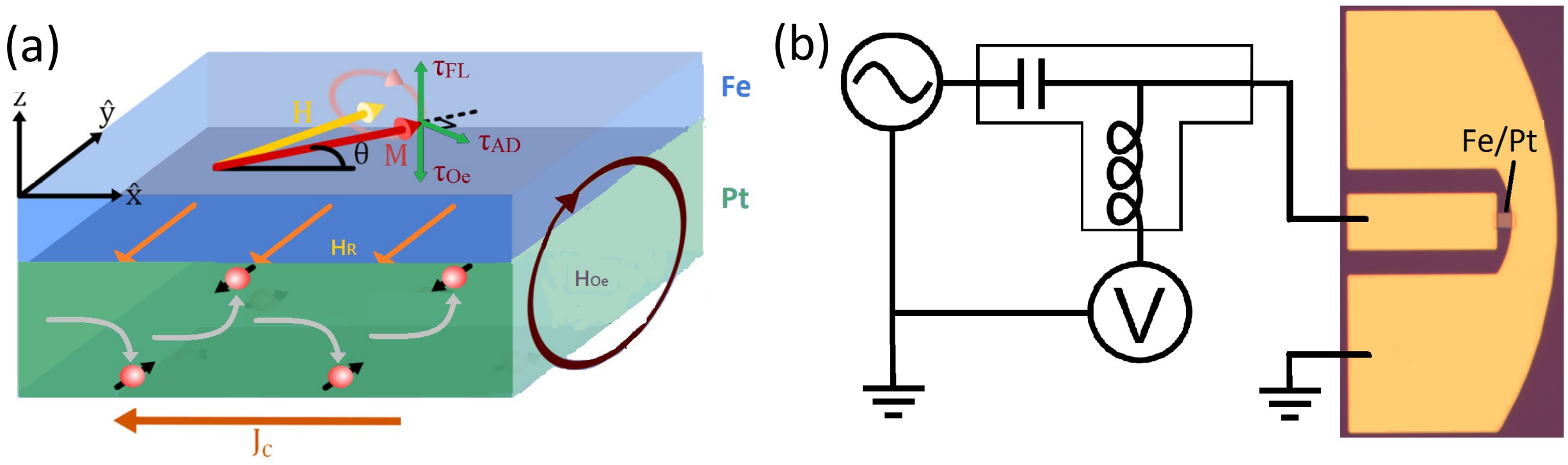

Figure 1.

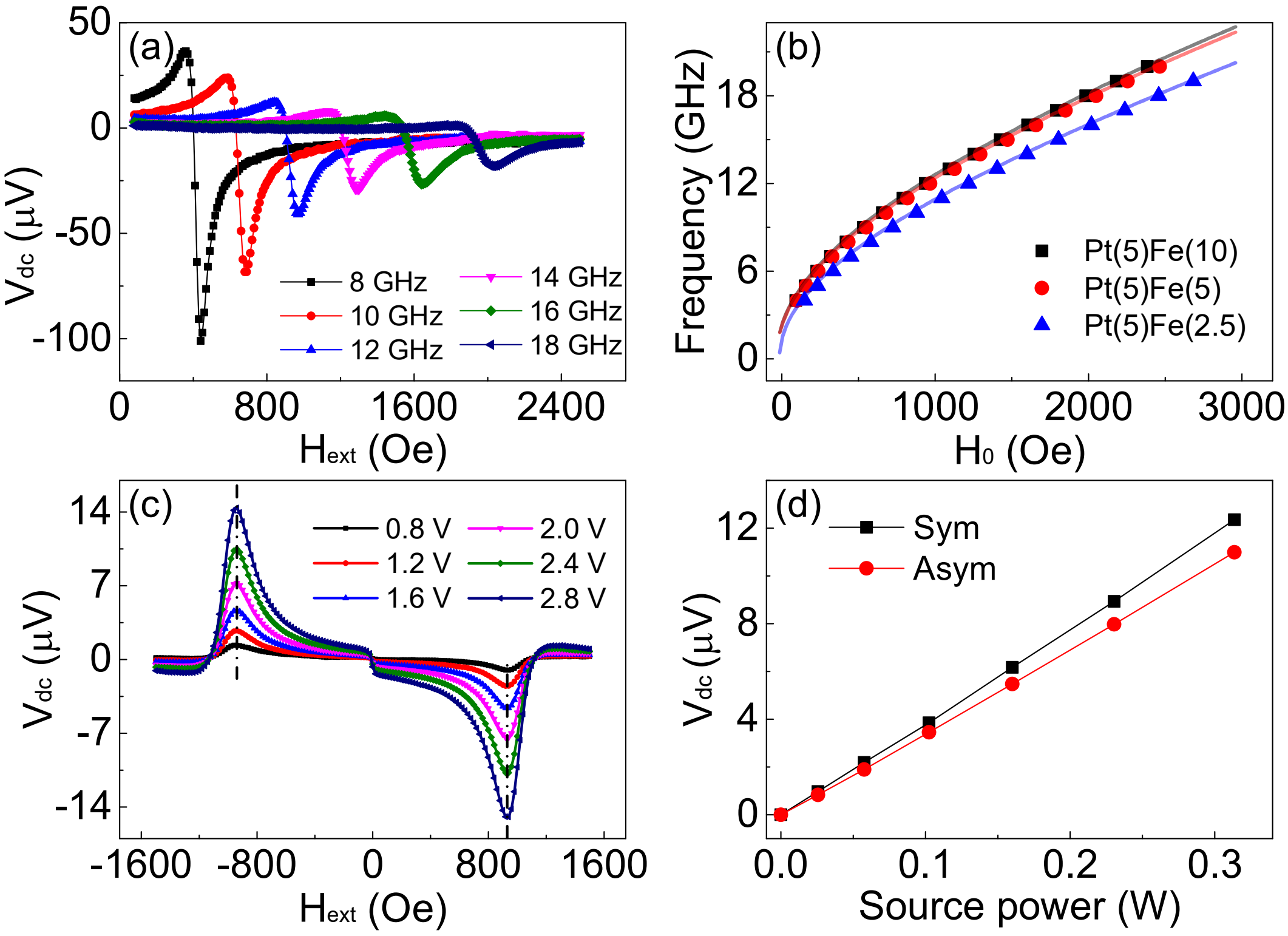

Figure 2.

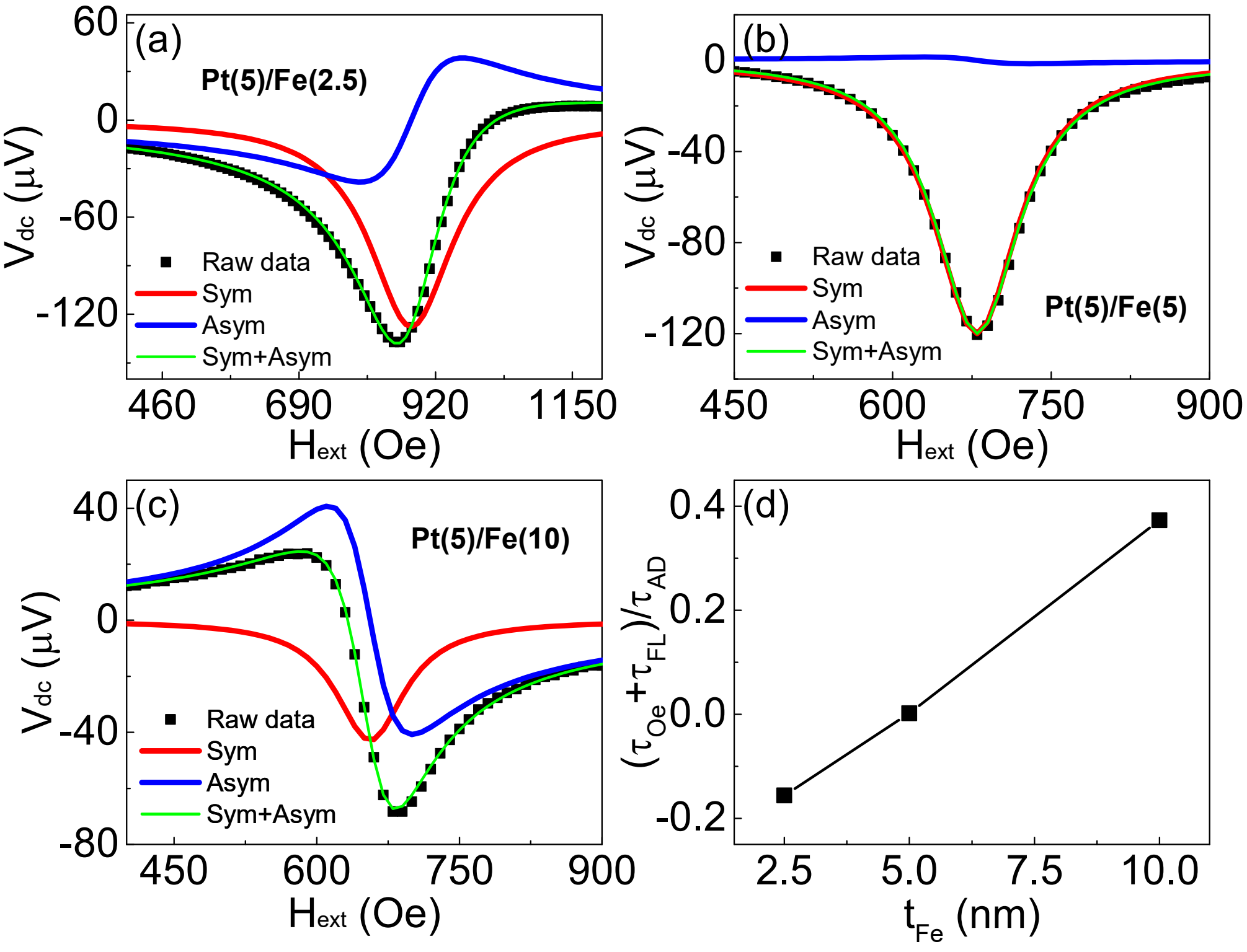

Figure 3.

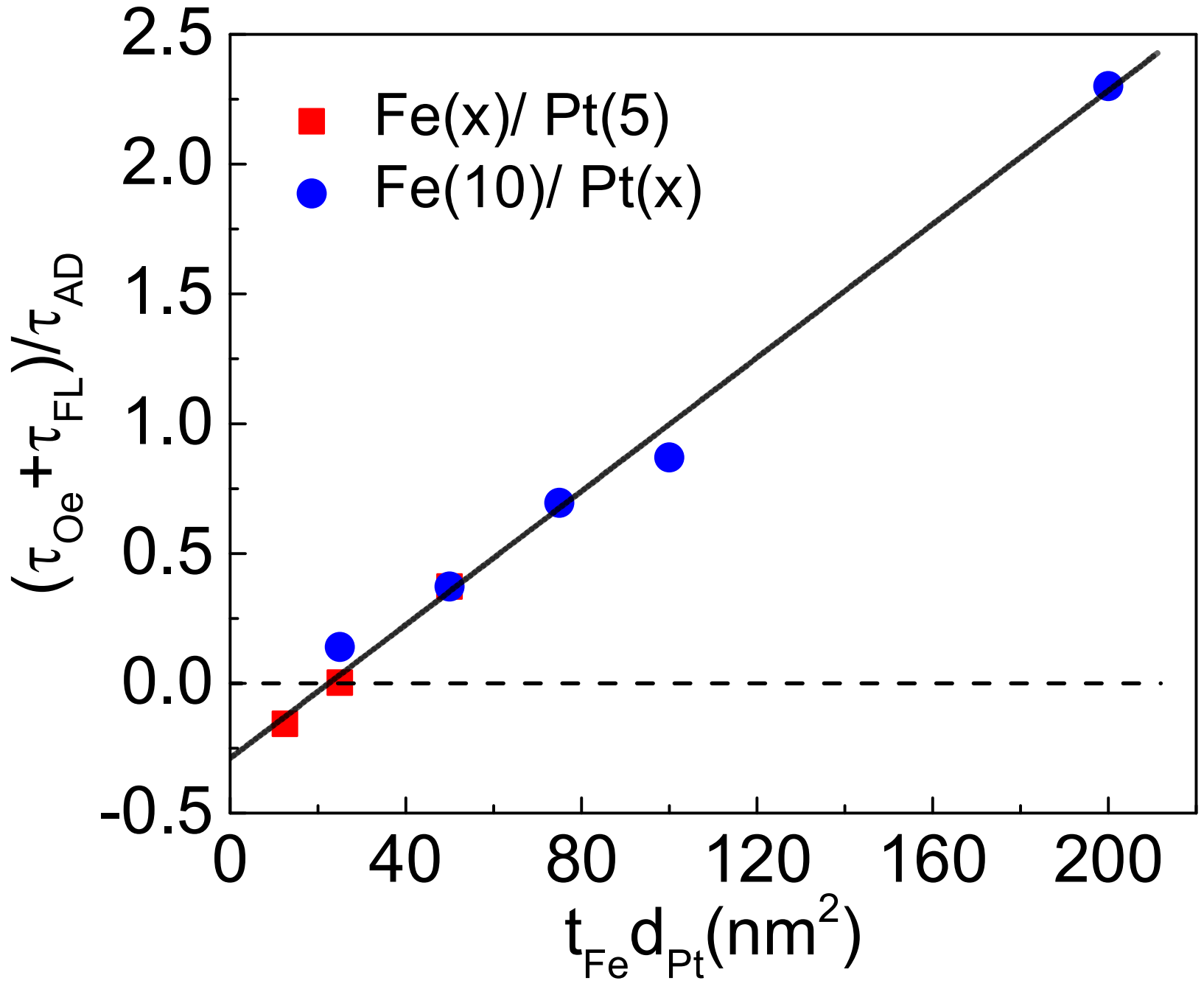

Figure 4.